\documentclass[5p,preprint]{elsarticle}
\usepackage{amsmath,amssymb,amsfonts,amsthm}
\usepackage{natbib}

\theoremstyle{plain}% Theorem-like structures provided by amsthm.sty
\usepackage{cuted}
\usepackage{float}
\newtheorem{theorem}{Theorem}[section]

\newtheorem{proposition}[theorem]{Proposition}
\theoremstyle{definition}
\newtheorem{definition}[theorem]{Definition}

\theoremstyle{remark}

\DeclareMathOperator*{\argmin}{arg\,min}
\DeclareMathOperator*{\trace}{tr}
\DeclareMathOperator*{\diag}{diag}

\usepackage[dvipsnames]{xcolor}

\usepackage[figuresright]{rotating}

\begin{document}

\begin{frontmatter}



%% e.g. \dochead{17th International Conference on Dynamical Processes in Excited States of Solids}

\title{Contraction Analysis and Control Synthesis for Discrete-time Nonlinear Processes}


\author{Lai Wei}
\author{Ryan McCloy} 
\author{Jie Bao\corref{cor1}}
\cortext[cor1]{Corresponding author: Jie Bao. This work was supported by ARC Discovery Grant DP210101978.}

\address{School of Chemical Engineering, The University of New South Wales, NSW 2052, Australia\\ (lai.wei1@unsw.edu.au, r.mccloy@unsw.edu.au, j.bao@unsw.edu.au)}

\begin{abstract}
Shifting away from the traditional mass production approach, the process industry is moving towards more agile, cost-effective and dynamic process operation (next-generation smart plants). This warrants the development of control systems for nonlinear chemical processes to be capable of tracking time-varying setpoints to produce products with different specifications as per market demand and deal with variations in the raw materials and utility (e.g., energy). This article presents a systematic approach to the implementation of contraction-based control for discrete-time nonlinear processes. Through the differential dynamic system framework, the contraction conditions to ensure the exponential convergence to feasible time-varying references are derived. The discrete-time differential dissipativity condition is further developed, which can be used for control designs for disturbance rejection. Computationally tractable equivalent conditions are then derived and additionally transformed into an SOS programming problem, such that a discrete-time control contraction metric and stabilising feedback controller can be jointly obtained. Synthesis and implementation details are provided and demonstrated through a numerical case study.
\end{abstract}

\begin{keyword}


discrete-time nonlinear processes \sep contraction theory \sep discrete-time control contraction metric (DCCM) \sep differential dissipativity \sep sum of squares (SOS) programming

\end{keyword}
\end{frontmatter}

% \linenumbers

\section{Introduction}
Chemical processes are traditionally designed for and operated at a certain steady-state operating condition, where the process economy is optimised. Nowadays, supply chains are increasingly dynamic and the process industry needs to shift from the traditional mass production to more agile, cost-effective and flexible process operations, in response to the fluctuations in market demand for products with different specifications, and the costs and supply of raw materials and energy.  As such, the control systems need to be able to drive the process safely and efficiently to any feasible time-varying operational targets, e.g., setpoints determined by the Real-time Optimisation (RTO) layer.

While most chemical processes are nonlinear, linear controllers have been commonly designed and implemented based on linearised models. This approach is defensible for regulatory control around a predetermined steady state. However, flexible process operation warrants nonlinear control as the target operating conditions can vary significantly to optimise economic costs (see, e.g., \cite{heidarinjad2021EMPC,kumar2002recycle}). Existing nonlinear control methods, e.g., Lyapunov-based approaches typically require redesigning the control Lyapunov function and controller when the target  equilibrium changes, not suitable for flexible process operation with time varying targets. To achieve time-varying setpoint tracking requires \emph{incremental stability} \cite{angeli2002lyapunov,santoso2012}.  This has motivated increased interest for alternative approaches based on the contraction theory framework \citep{WangBao17,mccloy2021differential, McCloyBao2021}. Introduced by \cite{lohmiller1998contraction}, contraction theory facilitates stability analysis and control of nonlinear systems with respect to arbitrary, time-varying (feasible) references without redesigning the control algorithm \citep{manchester2017control,lopez2019contraction}. Instead of using the state space process model alone, contraction theory also exploits the differential dynamics, a concept borrowed from fluid dynamics, to analyse the local stability of systems. Thus, one useful feature of contraction theory is that it can be used to analyse the incremental stability/contraction of nonlinear systems, and simultaneously synthesise a controller that ensures offset free tracking of feasible target trajectories using control contraction metrics (or CCMs, see, e.g., \cite{manchester2017control}). Furthermore, the dissipativity analysis  (\cite{willems2007dissipativity,ydstie2002passivity,garcia2016passivity}) was extended to differential system dynamics.  The resulting differential dissipativity \citep{forni2013diffdiss, WangBao17,mccloy2021differential} provides a framework to study the incremental input-output stability condition through consideration of differential storage functions and supply rates. Moreover, as virtually all modern process control systems are developed and implemented in a discrete time setting \citep{goodwin2001control}, there is a natural motivation for the development of tools for analysing, designing and implementing contraction-based control for discrete-time nonlinear systems. However, the current contraction-based (and more generally, differential system) analysis and control synthesis (e.g., \cite{manchester2014control,manchester2017control,manchester2018robust}), is limited to continuous-time control-affine nonlinear systems.

In the author's preliminary work \cite{wei2021control}, an SOS programming approach for the synthesis of contraction-based feedback control of discrete-time nonlinear systems was developed, utilising definitions for contracting systems in \cite{lohmiller1998contraction}. Leveraging the above results, we develop herein, the rigorous systematic analysis, control synthesis and implementation of contraction-based control through the use of SOS programming. Specifically, this article extends our previous work in \cite{wei2021control}, and provides a complete development and proofs for the exponential incremental stability (and convergence rates) of contracting discrete-time nonlinear processes, via discrete-time control contraction metrics (DCCMs). The above result has been further extended to  differential dissipativity conditions that ensure closed-loop differential $\mathcal{L}_2$ gain, which are useful in control design for disturbance attenuation. Tractable control synthesis approaches have been developed to achieve the desired contraction (for incremental stability) and dissipativity (for disturbance rejection) conditions. 

% Herein, this work provides complete development and proof for the incremental exponential stability properties and convergence rates of contracting discrete-time nonlinear processes via discrete-time control contraction metrics (DCCMs). Utilising the differential framework, dissipativity conditions resulting in closed-loop stability properties are also developed and shown to result in a bounded disturbance response. By leveraging the author's preliminary results in \cite{wei2021control}, systematic analysis, control synthesis and implementation of contraction-based control is presented through the use of SOS programming. 

The structure of this article is as follows. In Section~\ref{sec:DCCM diss}, the contraction and differential dissipativity analysis of control affine discrete-time nonlinear processes are presented, including the contraction and exponential differential dissipativity conditions. Section~\ref{sec:tractable} transforms the contraction and differential dissipativity conditions into computationally tractable synthesis problems. Based on these conditions,  Section~\ref{sec:synthesis} utilises the SOS programming to synthesise controllers, including details for controller implementation. Section~\ref{sec:example} illustrates the proposed contraction-based approach via a CSTR simulation study. Conclusions are drawn in Section~\ref{sec:conclusion}.  

\textbf{Notation.} Denote by $f_k = f(x_k)$ for any function $f$, $\mathbb{Z}$ represents the set of all integers, $\mathbb{Z}^+$ represents set of positive integers, $\mathbb{R}$ represents set of real numbers. $\Sigma$ represents a polynomial sum of squares function that is always non-negative, e.g. $\Sigma(x_k,u_k)$ is a polynomial sum of squares function of $x_k$ and $u_k$.

\section{Discrete-time Differential Dynamics based Convergence Analysis}\label{sec:DCCM diss}
In this section, we provide a complete treatment for the analysis and control of discrete-time control affine nonlinear systems using contraction theory. By extending to the controlled system setting through the use of discrete-time control contraction metrics (DCCMs), closed-loop stability of a contracting system is shown. Leveraging the differential framework, dissipativity properties are also assessed and translated into controller synthesis conditions. 

\subsection{Discrete-time Contraction Analysis}\label{sec:discrete-time contraction analysis}
\begin{figure*}[tb]
    \begin{center}
        \includegraphics[width=0.55\linewidth]{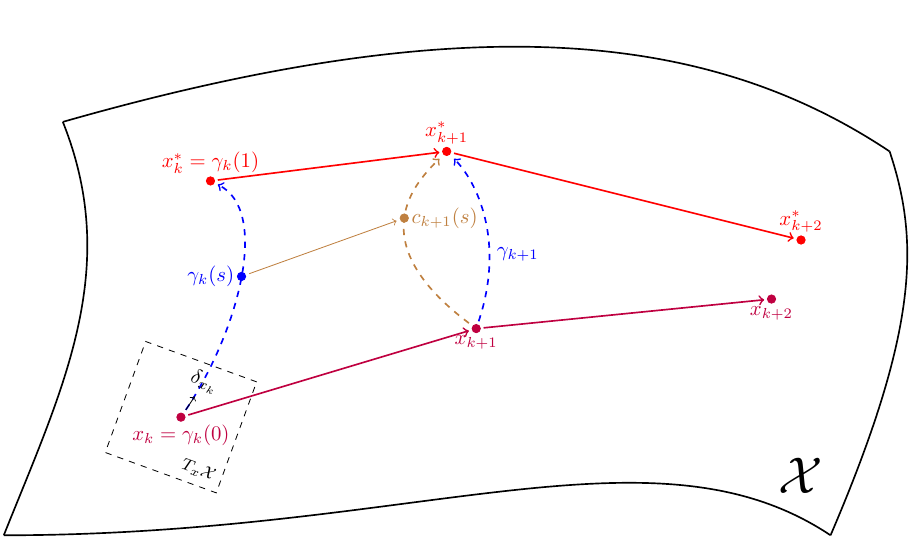}
        \caption{Illustration of the differential dynamics of a discrete-time system along the state manifold $\mathcal{X}$.}
        \label{fig:geodesic representation}
    \end{center}
\end{figure*}
In the following section, we introduce the necessary differential system framework and Riemannian tools underpinning contraction-based analysis and control approaches. 

To develop control approaches which are capable of delivering time-varying setpoints or reference profiles, we require reference-independent stability conditions, i.e., incremental stability (e.g., \cite{angeli2002lyapunov}). 
Contraction theory~\cite{lohmiller1998contraction,manchester2017control} uses the concept of displacement dynamics or \textit{differential dynamics} to assess incremental stability. Consider a discrete-time nonlinear system of the form,
\begin{equation}\label{equ:dynamical system}
    x_{k+1} = f(x_k),
\end{equation}
where $x_k \in \mathcal{X} \subseteq \mathbb{R}^n$ is state vector at time instant $k$, $\mathcal{X}$ represents “restricted” state space of $x_k$ for all $k$ and $f$ is a smooth differentiable function. 

Define the state trajectories of~\eqref{equ:dynamical system} as $\hat{x}=(x_0, \cdots, x_k, \cdots)$ for $k=0,\cdots,\infty$. The state trajectories can be parameterized by $s$ and thus described as $\hat{x}(s)$ $=$ $(x_1(s), x_2(s),$ $\cdots,$ $x_k(s),$ $\cdots)$, where $0 \leq s \leq 1$. For example, consider the trajectory $\hat{x}$, associated with $s=0$, and $\hat{x}^*$, associated with $s=1$, as shown in Figure~\ref{fig:geodesic representation}.
%Let the state trajectories of~\eqref{equ:dynamical system}, from $k=1,...,N$ by parameterized by $s$, e.g., $\tilde{x}(s) = x$  (see, e.g., Fig.~\ref{fig:geodesic representation}). 
Then, consider two neighboring discrete-time trajectories separated by an infinitesimal displacement $\delta_{x_k}$. Formally, $\delta_{x_k}:=\frac{\partial x_k}{\partial s}$ is a vector in the tangent space $T_x\mathcal{X}$ at $x_k$.
In this paper, the state manifold $\mathcal{X}$ and the tangent space $T_x\mathcal{X}$  can both be identified with $\mathbb{R}^n$.

The differential dynamics (or displacement dynamics) for system (\ref{equ:dynamical system}) are  defined as
\begin{equation}\label{equ:differential dynamical system}
    \delta_{x_{k+1}}= A_k\delta_{x_k},
\end{equation}where $A_k : = \frac{\partial f(x_k)}{\partial x_k}$ represents the Jacobian matrix of function $f$.
In the context of Riemannian geometry \cite{do1992riemannian}, a generalized infinitesimal displacement, $\delta_{z_k}$,  can be defined using a coordinate transformation described by a mapping from $\delta_{x_k}$ to $\delta_{z_k}$, using a square state-dependent matrix function $\Theta$, i.e.,
\begin{equation}\label{equ:coordinate transformation}
    \delta_{z_k} = \Theta(x_k)\delta_{x_k}.
\end{equation}
If we consider the evolution of the infinitesimal squared distance for system (\ref{equ:differential dynamical system}) at time step $k$ as $\delta_{x_k}^\top\delta_{x_k}$; then, naturally, the generalized squared distance for (\ref{equ:differential dynamical system}) can be defined using (\ref{equ:coordinate transformation}) as
%\begin{equation}\label{equ:generalised distance}
%    V(x_k,\delta_{x_k}) = \delta_{z_k}^\top\delta_{z_k} = \delta_{x_k}^\top\Theta(x_k)^\top\Theta(x_k)\delta_{x_k} =: \delta_{x_k}^\top M(x_k)\delta_{x_k},
%\end{equation}
\begin{equation}\label{equ:generalised distance}
\begin{aligned}
    V(x_k,\delta_{x_k}) &= \delta_{z_k}^\top\delta_{z_k} = \delta_{x_k}^\top\Theta(x_k)^\top\Theta(x_k)\delta_{x_k} \\ &=: \delta_{x_k}^\top M(x_k)\delta_{x_k},
\end{aligned}
\end{equation}
where the \textit{metric} $M(x_k) :=\Theta(x_k)^\top\Theta(x_k)$ is a symmetric positive-definite matrix function and is uniformly bounded, i.e., 
\begin{equation}\label{inequ:metric bound}
    \alpha_1 I \leq M(x_k) \leq \alpha_2 I \ \ \forall x_k \in \mathcal{X},
\end{equation}
for some positive constants $\alpha_2 \geq \alpha_1 > 0$. 
We then have the following definition for a contracting system:

\begin{definition}[\cite{lohmiller1998contraction}]
\label{def:contraction}
    System~\eqref{equ:dynamical system} is contracting, with respect to a uniformly bounded, positive definite metric, denoted by $M_k = M(x_k)$ and $M_{k+1} = M(x_{k+1})$, provided, $\forall x \in \mathcal{X}$ and $\forall \delta_x \in T_x\mathcal{X}$, 
    \begin{equation}\label{inequ:contraction condition}
        \begin{aligned}
            \delta^\top_{x_{k}} (A_k^\top M_{k+1} A_k - M_{k})\delta_{x_{k}}&\leq -\beta \delta_{x_k}^\top M_{k}\delta_{x_k} < 0,
        \end{aligned}
    \end{equation}
     for some constant $0 < \beta \leq 1$, where~\eqref{inequ:contraction condition} describes the region of contraction. In addition, the function $V_k$ in \eqref{equ:generalised distance} can be understood as a discrete-time Lyapunov function for the differential dynamics~\eqref{equ:differential dynamical system}, i.e., 
    \begin{equation}
    \label{eqe:Vdotcondition}
    V_{k+1} - V_k \leq - \beta V_k < 0, \quad \forall x \in \mathcal{X}, \forall \delta_x \in T_x\mathcal{X}.
    \end{equation}
\end{definition}
The contraction region defined in~\eqref{eqe:Vdotcondition} provides a generalized convergence condition for the discrete-time nonlinear system~\eqref{equ:dynamical system}, with respect to the metric $M_k$. Then, following some additional definitions from Riemannian geometry, incremental stability can be established.

For a smooth curve function, $c(s): [0,1] \rightarrow \mathcal{X}$ (see, Figure~\ref{fig:geodesic representation}), connecting any pair of points, $x,x^* \in \mathcal{X}$ (such that $c(0) = x$ and $c(1) = x^*$), we define the Riemannian distance, $d(x,x^*)$, and energy, $E(x,x^*)$, as (see, e.g., \cite{do1992riemannian})
\begin{equation}\label{equ:Riemannian distance and energy}
    \begin{aligned}
    d(x,x^*) = d(c) :=\int_0^1 \sqrt{\delta^\top_{c(s)}M(c(s))\delta_{c(s)}}ds,\\
    E(x,x^*) = E(c) :=\int_0^1\delta^\top_{c(s)}M(c(s))\delta_{c(s)}ds,
    \end{aligned}
\end{equation}
where $\delta_{c(s)} := \frac{\partial c(s)}{\partial s}$.
The shortest path in Riemannian space, or \textit{geodesic}, between $x$ and $x^*$ is defined as 
\begin{equation}\label{equ:geodesic}
    \gamma(s) :=\argmin_{c(s)} {d(x,x^*)}.
\end{equation}

\subsection{Discrete-time Control Contraction Metrics (DCCMs)}\label{sec:DCCM intro}
%Last several sections we discuss the discrete contraction theory and incremental stability, but it is not yet that useful to design a controller using these results. 
In this section, we complete an explicit proof for exponential incremental convergence analysis of contracting discrete-time nonlinear systems, and a controller structure for discrete-time contraction is developed (as an extension to existing \sloppy continuous-time results). We consider a discrete-time nonlinear system in control affine form as
\begin{equation}\label{equ:control affine}
    x_{k+1} = f(x_k) + g(x_k)u_k,
\end{equation}
where $x_k \in \mathcal{X} \subseteq \mathbb{R}^n$ and $u_k \in \mathcal{U} \subseteq \mathbb{R}^m$ are the state and control vectors, respectively, $\mathcal{X}$ and $\mathcal{U}$ represent the restricted state and control spaces, and, $f$ and $g$ are smooth functions.

We define a triplet solution sequence $(x_k^*,u_k^*,x_{k+1}^*)$ for system \eqref{equ:control affine} and consider the discrete-time control input, $u_k = u_k(x_k,x^*_k,u^*_k)$. The differential dynamics for system (\ref{equ:control affine}) satisfy \begin{equation}\label{equ:differential control affine}
    \delta_{x_{k+1}} = A(x_k)\delta_{x_k} + B(x_k)\delta_{u_k},
\end{equation} where, denoting $h(x_k,u_k) = f(x_k) + g(x_k)u_k$, the matrices $A:=\frac{\partial h(x_k,u_k)}{\partial x_k}$ and $B:=\frac{\partial h(x_k,u_k)}{\partial u_k}$ are the Jacobian matrices of functions $f$ and $g$ in \eqref{equ:control affine} respectively.   Consider a differential state-feedback control law \begin{equation}\label{equ:differential feedback}
    \delta_{u_k} = K(x_k) \delta_{x_k},
\end{equation} which leads to the following state-feedback law (by integrating \eqref{equ:differential feedback} over the geodesic $\gamma(x,x^*)$ \eqref{equ:geodesic}):
\begin{equation}\label{equ:control integral}
    u_k = u^*_k + \int_0^1K(\gamma(s))\frac{\partial \gamma(s)}{\partial s}\,ds.
\end{equation}
Note that this particular formulation is reference-independent, since the target trajectory variations do not require structural redesign of the feedback controller and is naturally befitting to the flexible manufacturing paradigm. Moreover, the discrete-time control input, $u_k$~\eqref{equ:control integral}, is a function with arguments ($x_k$, $x^*_k$, $u^*_k$) and hence the control action is computed from  the current state and target trajectory. 
\begin{theorem}
\label{theorem:incremental stability}
Consider a discrete-time nonlinear system~\eqref{equ:control affine} with differential dynamics \eqref{equ:differential control affine} and a differential state-feedback controller \eqref{equ:differential feedback} (implemented as a state-feedback controller \eqref{equ:control integral}). The above closed-loop system is contracting, with respect to a uniformly bounded, positive definite DCCM, $M(x_k)$, if
%A differential state-feedback control law is considered, given as $\delta_{u_k} = K(x_k)\delta_x_k$, the contraction condition in \eqref{inequ:contraction condition} can be formed.
\begin{equation}\label{inequ: control affine contraction condition}
    (A_k+B_kK_k)^\top M_{k+1}(A_k+B_kK_k) - (1-\beta)M_{k} < 0,
\end{equation}
for some constant $0 < \beta \leq 1$. 

Furthermore, for any feasible reference trajectory  $(\hat{x}^*, \hat{u}^*)$ (where $\hat{x}^*=(x^*_0, \cdots, x^*_k, \cdots)$ and $\hat{u}^*=(u^*_0, \cdots, u^*_k, \cdots)$) and an initial value  $x_0$ in the contraction region given by Definition  \ref{def:contraction}, the above closed-loop system is 
%\begin{enumerate}
    %\item 
    exponentially incrementally  stable, i.e.,
    \begin{equation}
    \label{equ:exp_stability}
        |x_k-x^*_k| \leq  R e^{-\lambda k \Delta_t} |x_0 - x^*_0|,
\end{equation}
for some constants $R$ and $\lambda$, where $x_k$ is the $k$-th step of the closed-loop state.
%\end{enumerate}
\end{theorem}
\begin{proof}
    Firstly, substituting \eqref{equ:differential feedback} into \eqref{equ:differential control affine} yields, the closed-loop differential system
    \begin{equation}\label{eq:diffcl}
        \delta_{x_{k+1}} = \left(A(x_k) + B(x_k)K(x_k)\right)\delta_{x_k}.
    \end{equation}
    Substituting the closed-loop differential dynamics \eqref{eq:diffcl} and generalised squared distance \eqref{equ:generalised distance} (as a differential Lyapunov candidate) into the differential Lyapunov condition of \eqref{eqe:Vdotcondition} provides the contraction condition in \eqref{inequ: control affine contraction condition} $\forall x ,\delta_x \in \mathcal{X}$.
    Then, consider $(x_k,x_{k+1})$ and $(x^*_{k},x^*_{k+1})$ as two solution pairs of system~\eqref{equ:dynamical system}. The shortest path connecting each of these solutions at time step $k$, denoted by the geodesic $\gamma_k(s)$ in~\eqref{equ:geodesic}, is defined with $\gamma_k(0) = x_k$ and $\gamma_k(1) = x_k^*$, and $c_{k+1}(s)$ denotes a path (not necessarily a geodesic) at the next time step, connecting $c_{k+1}(0) = x_{k+1}$ and $c_{k+1}(1)=x_{k+1}^*$. Hence, from \eqref{equ:dynamical system}, we have $c_{k+1}(0) = x_{k+1} = f(x_k) =f(\gamma_k(0))$ and $c_{k+1}(1) = x_{k+1}^* = f(x_k^*) = f(\gamma_k(1))$ as shown in Figure~\ref{fig:geodesic representation},
    or, without loss of generality,
    \begin{equation}\label{equ:geodesic k to path k+1}
        \begin{aligned}
            c_{k+1}(s) = f(\gamma_k(s)).
        \end{aligned}
    \end{equation}
    From \eqref{equ:Riemannian distance and energy}, \eqref{equ:geodesic} and the Hopf-Rinow Theorem, (e.g., \cite{manchester2017control}), we have
    \begin{equation}\label{equ:Hopf-Rinow}
        E(\gamma) = d(\gamma)^2 \leq d(c)^2 \leq E(c),
    \end{equation}
    which, given \eqref{inequ:contraction condition} and \eqref{equ:geodesic k to path k+1}, yields decreasing Riemannian energy, i.e.,
    \begin{equation}{\label{equ:energy bound}}
        \begin{aligned}
        E(\gamma_{k+1}) 
        & \leq \int_0^1 {\frac{\partial c_{k+1}(s)}{\partial s}}^\top M(c_{k+1}(s))\frac{\partial c_{k+1}(s)}{\partial s} ds \\
        & \leq \int_0^1 (1-\beta) {\frac{\partial \gamma_k(s)}{\partial s}}^\top M(\gamma_k(s))\frac{\partial \gamma_k(s)}{\partial s} \\
        &= (1-\beta)E(\gamma_k).
        \end{aligned}
    \end{equation}
    Taking the square root of \eqref{equ:energy bound} gives
    \begin{equation}\label{inequ:distance bound}
        d(\gamma_{k+1}) \leq (1-\beta)^{\frac{1}{2}} d(\gamma_k).
    \end{equation}
    Since $M_k$ is by definition uniformly bounded (by two positive constants $\alpha_1$ and $\alpha_2$), then from \eqref{equ:Riemannian distance and energy} and \eqref{equ:geodesic}, we have
    \begin{equation}\label{inequ:a1 a2 bound}
        \sqrt{\alpha_1}|x(k)-x^*(k)| \leq d(\gamma(k)),\ \ 
        d(\gamma(0)) \leq \sqrt{\alpha_2}|x(0)-x^*(0)|.
    \end{equation}
    Combining \eqref{inequ:distance bound} and \eqref{inequ:a1 a2 bound} we have 
    \begin{equation}
        |x_k-x^*_k| \leq (1-\beta)^{\frac{k}{2}} \sqrt{\frac{\alpha_2}{\alpha_1}} |x_0 - x^*_0|,
    \end{equation}
    which for $R = \sqrt{\frac{\alpha_2}{\alpha_1}} $ yields \begin{equation}
    \label{equ:asym_stability}
        |x_k-x^*_k| \leq (1-\beta)^{\frac{k}{2}} R|x_0 - x^*_0|.
    \end{equation}
    Since the contraction rate, $\beta$, is by definition bounded as $ 0 < \beta \leq 1$, let $e^{-\lambda k \Delta_t} = (1-\beta)^{\frac{k}{2}}$ for some positive constant $\lambda$. Furthermore, we can express \eqref{inequ:distance bound} with respect to a discrete-time interval, $\Delta_t$, i.e.,  
    \begin{equation}
    \label{equ:asy2exp}
        d(\gamma_{k\Delta_t}) \leq (1-\beta)^{\frac{k}{2}} d(\gamma_0) = e^{-\lambda k \Delta_t} d(\gamma_0).
    \end{equation}
    Then, provided the $\lambda$ satisfying \eqref{equ:asy2exp}, e.g., for $\lambda =-\ln(1-\beta)/(2\Delta_t)$, we have the result in \eqref{equ:exp_stability} by substituting \eqref{equ:asy2exp} into \eqref{inequ:a1 a2 bound}.

Now, we consider the controlled system. The contraction condition in \eqref{inequ: control affine contraction condition} is obtained by substituting the corresponding differential dynamic equation \eqref{equ:differential control affine} and differential feedback control law \eqref{equ:differential feedback} into the autonomous contraction condition \eqref{inequ:contraction condition}. The control law in \eqref{equ:control integral} can be obtained by integrating the differential controller (\ref{equ:differential feedback}) along the geodesic, with respect to metric, $M_k$, in \eqref{equ:generalised distance} and the stability result is straightforward from the above proof by replacing the contraction condition \eqref{inequ:contraction condition} by \eqref{inequ: control affine contraction condition}.
\end{proof}

To summarize, a suitably designed contraction-based controller ensures that the length of the minimum path (i.e., geodesic) between any two trajectories (e.g., the plant state, $x$, and desired state, $x^*$, trajectories), with respect to the metric $M$, shrinks with time, i.e., provided that the contraction condition \eqref{inequ: control affine contraction condition} holds for the discrete-time nonlinear system \eqref{equ:control affine}, %\LAI{in the contraction region}, 
we can employ a stabilizing feedback controller \eqref{equ:control integral} to ensure convergence to feasible operating targets.

\subsection{Exponential Differential Dissipativity}
% ext. above contraction.. to diff diss. eq 30-32+ where dist rej is an app. of such framework.
In the following, we extend differential dissipativity results \cite{WangBao17} to the discrete-time setting and use this property to determine bounded disturbance characteristics of the corresponding nonlinear system. Consider the discrete-time nonlinear system with disturbance, $\nu$, in the form of
\begin{equation}\label{eq:sys_d}
x_{k+1} = f(x_k)x_k + g_u(x_k)u_k + g_{\nu}(x_k)\nu_k,
\end{equation}
with corresponding differential dynamics described by \begin{equation}\label{eq:sys_d diff}
    \delta_{x_{k+1}} = A\delta_{x_k} +  B_{u,k}\delta_{u_k} +  B_{\nu,k}\delta_{\nu_k}, 
\end{equation}
where $A = \frac{\partial h(x_k,u_k,\nu_k)}{\partial x_k}$, $B_u = \frac{\partial h(x_k,u_k,\nu_k)}{\partial u_k}$ and $B_\nu = \frac{\partial h(x_k,u_k,\nu_k)}{\partial \nu_k}$ and $h(x_k,u_k,\nu_k) = f(x_k)x_k + g_u(x_k)u_k + g_{\nu}(x_k)\nu_k$. 
\begin{definition}\label{def:diff diss}
System \eqref{eq:sys_d}--\eqref{eq:sys_d diff} is exponentially differentially dissipative with respect to the differential supply rate $\sigma(\delta_x,\delta_d)$, if there exists a differential storage function $V(x,\delta_x)$ such that
\begin{equation}\label{eq:diff_diss}
    V_{k+1} - (1 - \beta) V_k\leq \sigma_k.
\end{equation}
Moreover, the system is called exponentially differentially \sloppy (Q,S,R)-dissipative if \eqref{eq:diff_diss} is satisfied with respect to the partitioned supply rate
\begin{equation}\label{eq:qsr def}
    \sigma = 
\begin{bmatrix}
\delta_x \\ \delta_\nu
\end{bmatrix}^\top
\begin{bmatrix} Q & S\\S^\top & R \end{bmatrix}
\begin{bmatrix}
\delta_x \\ \delta_\nu
\end{bmatrix}.
\end{equation}
\end{definition}
 
One particular application for the differential dissipativity framework lies in systematically determining physical properties of the closed-loop system. For example, satisfying \eqref{eq:diff_diss} with $Q=0$, $S=0.5I$, $R=0$, indicates the system is differentially passive. Furthermore, differential dissipativity with respect to a supply rate $Q<0$, $S=0$, $R>0$ implies a differential $\mathcal{L}_2$ gain bound. This differential dissipativity condition can  be generalized to determine the incremental stability of the corresponding nonlinear closed-loop system (e.g., the $\mathcal{L}_2$ gain from the disturbances to states), which is detailed below.
\begin{theorem}\label{thm:dif dis}
System \eqref{eq:sys_d},\eqref{eq:sys_d diff},\eqref{equ:control integral} is exponentially differentially (Q,S,R)-dissipative, provided a DCCM and differential feedback gain pair, $(M,K)$, are found satisfying
\begin{equation}\label{equ:diff diss sub}
\begin{split}
    &\delta_{x_{k}}^\top(A_k+B_{u,k} K_k)^\top M_{k+1}(A_k+B_{u,k} K_k)\delta_{x_{k}} - (1-\beta)\delta_{x_{k}}^\top M_{k} \delta_{x_{k}} \\ 
    &\quad + \delta^\top_{\nu_k} B^\top_{\nu_k}M_{k+1}B_{\nu_k}\delta_{\nu_k}  \leq \delta_{x_{k}}^\top Q \delta_{x_{k}} + 2 \delta_{\nu_{k}}^\top S \delta_{x_{k}} + \delta_{\nu_{k}}^\top R \delta_{\nu_{k}}.
\end{split}
%\delta_{x_{k+1}}^\top M_{k+1} \delta_{x_{k+1}} - (I - \beta)\delta_{x_k}^\top M_{k} \delta_{x_k} \leq
%  \begin{bmatrix}\delta_{x_k}\\ \delta_\nu\end{bmatrix}^\top \begin{bmatrix} Q &\! S\\S^\top & \!R \end{bmatrix}
% \begin{bmatrix}\delta_{x_k}\\ \delta_\nu\end{bmatrix}. 
\end{equation}
Furthermore, the closed-loop nonlinear system, \eqref{eq:sys_d} driven by \eqref{equ:control integral} with $Q<0$ and $R>0$ satisfying \eqref{equ:diff diss sub}, is incrementally exponentially stable, and exhibits a bounded disturbance response, for any feasible  reference $x_k^*$ and expected disturbance $\nu_k^*$ (typically zero for unknown disturbances). Over the finite interval $k \in [0,N]$,
\begin{equation}\label{eq:bounded dist response}
    \sum_{k=0}^N ||x_k - x_k^* ||_2 \leq \rho \sum_{k=0}^N||\nu_k - \nu_k^* ||_2,
\end{equation} 
where $\rho$ is the incremental truncated $\mathcal{L}_2$ gain from $\nu$ to $x$.
\end{theorem}
\begin{proof}
Substituting \eqref{equ:differential feedback} into \eqref{eq:sys_d diff} yields, the closed-loop differential system
    \begin{equation}\label{eq:diffcl_d}
        \delta_{x_{k+1}} = \left(A(x_k) + B_u(x_k)K(x_k)\right)\delta_{x_k} + B_\nu(x_k)\delta_{\nu_k}.
    \end{equation}
    Given a differential $(Q,S,R)$-supply rate as in \eqref{eq:qsr def} and substituting the closed-loop differential dynamics \eqref{eq:diffcl_d} and generalised squared distance \eqref{equ:generalised distance} (as a differential storage function) into the exponential differential dissipativity condition of \eqref{eqe:Vdotcondition} provides the  exponential differential $(Q,S,R)$-dissipativity condition in \eqref{equ:diff diss sub}.
% Choosing the differential storage function as the infinitesimal squared distance \eqref{equ:generalised distance}%$V_k = \delta_{x_k} M_k \delta_{x_k}$
% , given a differential $(Q,S,R)$-supply rate as in \eqref{eq:qsr def},
% and substituting in the differential dynamics \eqref{eq:sys diff} with differential feedback controller \eqref{equ:differential feedback} into \eqref{eq:diff_diss}, gives the exponential differential $(Q,S,R)$-dissipativity condition in \eqref{equ:diff diss sub}.
Exponential incremental stability for \eqref{eq:sys_d} when $Q<0$, is then straightforward from the proof of Theorem \ref{theorem:incremental stability} by integrating \eqref{equ:diff diss sub} along the geodesic $\gamma(x,x^*)$ \eqref{equ:geodesic}. 

Moreover, to see the that the disturbance response of \eqref{eq:sys_d} is bounded on the finite interval $k\in[0,N]$, integrate the differential dissipativity condition \eqref{eq:diff_diss} (omitting the $\beta V_k$ term, since the rate of change in the differential storage function is not relevant to the incremental $\mathcal{L}_2$ gain) along the geodesic $\gamma(x,x^*)$ and sum over all $k \in [0,N]$ to obtain
\begin{equation}\label{eq:sum_int_bounded}
    \sum_{k=0}^N \int_0^1 (V_{k+1} - V_k )ds \leq \sum_{k=0}^N \int_0^1 \left( \delta_x^\top Q \delta_x + 2\delta_x^\top S \delta_\nu + \delta_\nu^\top R \delta_\nu \right)ds.
\end{equation}
By induction and definition for Riemannian energy, \eqref{eq:sum_int_bounded} yields
\begin{equation}
    E(\gamma_{N+1}) - E(\gamma_0) \leq \sum_{k=0}^N \int_0^1 \delta_x^\top Q \delta_x + 2\delta_x^\top S \delta_\nu + \delta_\nu^\top R \delta_\nu.
\end{equation}
Given an equivalent initial condition $E(\gamma_0) = 0$ (i.e., $x(0)=x^*(0)$ -- reasonable via definition of the initial time step $k=0$), and noting that by definition $E(\gamma) \geq 0$, gives
\begin{equation}
   \sum_{k=0}^N \int_0^1  \delta_x^\top Q \delta_x + 2\delta_x^\top S \delta_\nu + \delta_\nu^\top R \delta_\nu \, \,  ds \geq E(\gamma_{N+1}) \geq 0.
\end{equation}
Denoting $\hat{Q}  = -Q$ and completing the square gives 
\begin{equation}
    \sum_{k=0}^N \int_0^1 \delta_\nu^\top (R+S^\top \hat{Q}^{-1}S)\delta_\nu - ||\hat{Q}^{\frac{1}{2}} \delta_x - \hat{Q}^{-\frac{1}{2}} S \delta_\nu ||^2_2 \,\,ds \geq 0.
\end{equation}
Evaluating the integral by the Cauchy-Schwarz inequality, parameterising the disturbance input as $\nu_k(s) = (1-s)\nu^*_k +s\nu_k$ and noting that $d_\gamma(x,x^*) \leq ||x-x^*||$ 
gives
\begin{equation}
  \sum_{k=0}^N (R+S^\top \hat{Q}^{-1}S) ||\nu_k - \nu_k^* ||^2_2 -  ||\hat{Q}^{\frac{1}{2}}(x_k - x_k^*) - \hat{Q}^{-\frac{1}{2}} S (\nu_k - \nu_k^*)||^2_2 \geq 0,
\end{equation}
 and hence \eqref{eq:bounded dist response},
% \begin{equation}
%     \sum_{k=0}^N ||x_k - x_k^* ||_2 \leq \rho \sum_{k=0}^N||\nu_k - \nu_k^* ||_2,
% \end{equation} 
where $\rho = ||\hat{Q}^{-\frac{1}{2}}||_2(||\hat{Q}^{-\frac{1}{2}} S||_2 + (R+S^\top \hat{Q}^{-1}S)^{\frac{1}{2}})$. %is  the incremental truncated $L_2$ gain from $d$ to $x$.
%Hence the state response to the disturbance is bounded. 
\end{proof}

Note that these dissipativity results can be further extended to output feedback systems, or tailored for performance measures that are governed by particular states or outputs, by considering a measured output, e.g., $y_k = C(x_k)x_k$, and suitably modifying the dissipativity condition in \eqref{eq:diff_diss} (with arguments $(y,\delta_y)$). The disturbance response with respect to output $y$, could then be shaped by a weighted $\mathcal{L}_2$ gain from $d$ to $y$. 

Note that the $\beta V_k$ term is not relevant to the incremental $\mathcal{L}_2$ gain (see the proof of Theorem \ref{thm:dif dis}). However, this term ensures the exponential incremental stability when the disturbance is zero (hence the results of Theorem \ref{thm:dif dis} are permitted to collapse to those of Theorem \ref{theorem:incremental stability}). 

It is also interesting to note that the dissipativity condition in \eqref{equ:diff diss sub} links the input to the contraction condition in \eqref{inequ: control affine contraction condition}, whereby satisfaction of \eqref{equ:diff diss sub} ensures additional closed-loop properties, by imposing additional conditions on the permitted controller. As an example, by choosing $Q=-I$, $S=0$ and $R =  \alpha^2 I$, when solving for the pair $(M,K)$ in \eqref{equ:diff diss sub}, yields an incremental truncated $\mathcal{L}_2$ gain equal to $\alpha$, providing an additional design variable (implicit to the contraction-based controller \eqref{equ:control integral}) with respect to disturbance rejection. 

%\RMA{Perhaps this leads to considering saturation of the actuator? (considered as a disturbance effectively?)}

\section{Tractable Contraction Metric and Controller Conditions for Discrete-time Nonlinear Systems}\label{sec:tractable}

This section presents the transformation of contraction and differential dissipativity conditions (developed in Section \ref{sec:DCCM diss}) for discrete-time control-affine nonlinear systems into a tractable synthesis problem. 
\subsection{Obtaining a Tractable Contraction Condition}
As characterised by Theorem \ref{theorem:incremental stability}, two conditions are needed to ensure contraction of discrete-time nonlinear systems -- the first is the discrete-time contraction condition \eqref{inequ: control affine contraction condition}, and the second is the positive definite property of the metric $M$. Inspired by \cite{manchester2017control}, an equivalent condition to \eqref{inequ: control affine contraction condition} is developed in the following Proposition as a computationally tractable means for handling the coupled terms in \eqref{inequ: control affine contraction condition}. 

\begin{proposition}\label{thm:condition}
    Consider a differential feedback controller \eqref{equ:differential feedback} for the differential dynamics \eqref{equ:differential control affine} of a discrete-time control-affine nonlinear system \eqref{equ:control affine}. The discrete-time nonlinear system \eqref{equ:control affine} is contracting with respect to a DCCM, $M$, if a pair of matrix functions $(W,L)$ satisfies 
    \begin{equation}\label{inequ:LMI}
        \begin{bmatrix}
            W_{k+1}              & A_kW_k+B_kL_k \\
            (A_kW_k+B_kL_k)^\top    & (1-\beta)W_k
        \end{bmatrix} > 0,
    \end{equation}
    where $A_k,B_k$ and $K_k$ are functions in \eqref{equ:control affine} and \eqref{equ:differential feedback} respectively, $W_{k}:=M_k^{-1}$, $W_{k+1}:= M_{k+1}^{-1}= M^{-1}(f(x_k)+B(x_k)u_k)$, $L_k := K_kW_k$ and $\beta \in (0,1]$.
\end{proposition}
\begin{proof}
    Condition \eqref{inequ: control affine contraction condition} is equivalent to 
    \begin{equation}\label{Metric_Condi}
        (1-\beta)M_{k} - (A_k+B_kK_k)^\top M_{k+1}(A_k+B_kK_k) > 0.
    \end{equation}
    Applying Schur's complement \cite{boyd1994linear} to (\ref{Metric_Condi}) yields
    \begin{equation}
    \label{equ:schursCC}
        \begin{bmatrix}
            M_{k+1}^{-1} & (A_k+B_kK_k) \\
            (A_k+B_kK_k)^\top     & (1-\beta)M_k
        \end{bmatrix} > 0.
    \end{equation}
    Defining $W_k := M^{-1}(x_k)$ and $W_{k+1} := M^{-1}(x_{k+1}) = M^{-1}(f(x_k)+B(x_k)u_k)$, we then have
    \begin{equation}
    \label{equ:inequality_step}
        \begin{bmatrix}
            W_{k+1}     & (A_k+B_kK_k) \\
            (A_k+B_kK_k)^\top    & (1-\beta)W_k^{-1}
        \end{bmatrix} > 0.
    \end{equation}
    Left/right multiplying \eqref{equ:inequality_step} by an invertible positive definite matrix, $\diag\{I,W_k\}$ (and its transpose), yields
    % \begin{equation}
    %     \begin{bmatrix}
    %         I   && 0 \\
    %         0   && W_k
    %     \end{bmatrix}^\top
    %     \begin{bmatrix}
    %         W_{k+1}     && (A_k+B_kK_k) \\
    %         (A_k+B_kK_k)^\top    && (1-\beta)W_k^{-1}
    %     \end{bmatrix}
    %     \begin{bmatrix}
    %         I   && 0 \\
    %         0   && W_k
    %     \end{bmatrix}  > 0,
    % \end{equation}
    % which is equivalent to the following condition
    \begin{equation}
        \begin{bmatrix}
            W_{k+1}          & (A_k+B_kK_k)W_k \\
            W_k^\top (A_k+B_kK_k)^\top     & (1-\beta)W_k
        \end{bmatrix} > 0.
    \end{equation}
    Finally, defining $L_k:=K_kW_k$, we have the condition \eqref{inequ:LMI}. 
\end{proof}

\subsection{Exponential Differential Dissipativity Conditions}\label{sec:dissipativity}
As described in Theorem \ref{thm:dif dis}, dissipativity properties of the closed-loop can be formed by considering a differential storage function and supply rate, yet the resulting condition \eqref{equ:diff diss sub} is computationally complex. Even with the supply rate  provided, finding a DCCM and feedback gain satisfying \eqref{equ:diff diss sub} is at least as difficult (e.g., considering coupled terms) as the ``reduced'' inequality \eqref{inequ: control affine contraction condition} (forming the left-hand side of \eqref{equ:diff diss sub}). 
Inspired by \cite{wang2017distributed}, an equivalent computationally tractable condition to \eqref{equ:diff diss sub} is developed in the following Proposition. 
%\RMA{
% Consider
% \begin{itemize}
%     \item eq 38 
%     \item eq 42 (potentially incorrect, but most attractive)
%     \item eq 44
%     \item eq 48
% \end{itemize}}
\begin{proposition}\label{prop:dif dis}
 Consider a differential feedback controller \eqref{equ:differential feedback} for the differential dynamics \eqref{eq:sys_d diff} of a discrete-time control-affine nonlinear system \eqref{eq:sys_d}. The system is exponentially incrementally differentially (Q,S,R)-dissipative (in the sense of Definition \ref{def:diff diss}), with respect to a DCCM, $M$, and differential (Q,S,R,)-supply rate \eqref{eq:qsr def}, if a pair of matrix functions $(W,L)$ satisfies 
\begin{equation}\label{equ:LMI_dd}
% \begin{bmatrix} 
% W_{k+1} & A W_k + B_c L & B_\nu & 0\\
% (A W_k + B_c L)^\top & (I - \beta)W_k & W_k S & W_k \\ 
% B_\nu^\top & S^\top W_k & R & 0\\
% 0 & W_k & 0 & -Q^{-1} 
% \end{bmatrix}  \geq 0.
\begin{bmatrix} 
W_{k+1} & A_k W_k + B_{u,k} L_k & B_{\nu,k} & 0\\
(A_k W_k + B_{u,k} L_k)^\top & (1 - \beta)W_k & W_k S & W_k \\ 
B_{\nu,k}^\top & S^\top W_k & R & 0\\
0 & W_k & 0 & -Q^{-1} 
\end{bmatrix}  \geq 0,
\end{equation}
where $A_k$, $B_{u,k}$, $B_{\nu,k}$ and $K_k$ are defined in \eqref{eq:sys_d diff} and \eqref{equ:differential feedback} respectively, $W_{k}:=M_k^{-1}$, $W_{k+1}:= M_{k+1}^{-1}= M^{-1}(f(x_k)+B(x_k)u_k)$, $L_k := K_kW_k$ and $\beta \in (0,1]$.
\end{proposition}
\begin{proof}
From \eqref{equ:diff diss sub}, we have
\begin{equation}
\begin{split}
\begin{bmatrix}\delta_{x}\\ \delta_\nu\end{bmatrix}^\top 
\begin{bmatrix}A_{c,k} & B_{\nu,k} \end{bmatrix}^\top M_{k+1} \begin{bmatrix}A_{c,k} & B_{\nu,k} \end{bmatrix}
\begin{bmatrix}\delta_{x}\\ \delta_\nu\end{bmatrix}
-  (1 - \beta) \delta_x^\top M_k \delta_x 
 \\ \leq
\begin{bmatrix}\delta_{x}\\ \delta_\nu\end{bmatrix}^\top \begin{bmatrix} Q & S\\S^\top & R \end{bmatrix}
\begin{bmatrix}\delta_{x}\\ \delta_\nu\end{bmatrix}, 
\end{split}
\end{equation}
where $B_{\nu,k} = A_k + B_{u,k} K_k$. This inequality can be equivalently written
\begin{equation}
\begin{bmatrix} Q + (1 - \beta)M_k & S\\S^\top & R \end{bmatrix} -
\begin{bmatrix}A_{c,k} & B_{\nu,k} \end{bmatrix}^\top M_{k+1} \begin{bmatrix}A_{c,k} & B_{\nu,k} \end{bmatrix}  \geq 0.
\end{equation}
By Schur's complement, this is equivalent to
\begin{equation}
\begin{bmatrix} 
M_{k+1}^{-1} & A_{c,k} & B_{\nu,k} \\
A_{c,k}^\top & Q + (1 - \beta)M_k & S\\ 
B_{\nu,k}^\top & S^\top & R 
\end{bmatrix}  \geq 0.
\end{equation}
% Multiplying through by $diag\{M_+,I\}$ gives
% \begin{equation}
% \begin{bmatrix} 
% M_+^\top & M_+^\top A_c & M_+^\top B_d \\
% A_c^\top M_+ & Q + (I - \beta)M & S\\ 
% B_d^\top M_+ & S^\top & R
% \end{bmatrix} \geq 0,
% \end{equation}
% and defining $Y=M_+ B_c K$ (requires online $K = (B_c^\top B_c)^{-1} B^\top M_+^{-1} Y$) gives
% \begin{equation}
% \begin{bmatrix} 
% M_+^\top & M_+^\top A + Y & M_+^\top B_d \\
% A^\top M_+ + Y^\top & Q + (I - \beta)M & S\\ 
% B_d^\top M_+ & S^\top & R
% \end{bmatrix} \geq 0,
% \end{equation}
Defining $W_k := M^{-1}_k$ %gives
% \begin{equation}
% \begin{bmatrix} 
%  W_{k+1}+ & A_c & B_\nu  \\
%  A_c^\top & Q + (I - \beta)W_k^{-1} & S\\ 
%  B_\nu^\top & S^\top & R
% \end{bmatrix}  \geq 0
% \end{equation}
and appropriately (left/right) multiplying by $\diag\{ I , W_k, I\}$ (and its transpose) gives
\begin{equation}
\begin{bmatrix} 
W_{k+1}  & A_{c,k} W_k & B_{\nu,k} \\
W_k^\top A_{c,k}^\top & W_k^\top Q W_k + (1 - \beta)W_k^\top & W_k^\top S\\ 
B_{\nu,k}^\top & S^\top W_k & R
\end{bmatrix}  \geq 0.
\end{equation}
Applying Schur's complement, expanding $A_{c,k}$, %this is equivalent to
% \begin{equation}
% \begin{bmatrix} 
% W_{k+1} & A_c W_k & B_\nu & 0\\
% W_k^\top A_c^\top & (I - \beta)W_k^\top & W_k^\top S & W_k^\top \\ 
% B_\nu^\top & S^\top W_k & R & 0\\
% 0 & W_k & 0 & -Q^{-1} 
% \end{bmatrix}  \geq 0.
% \end{equation}
defining $L_k := K_k W_k$ and noting by symmetry $W_k^\top = W_k$ gives \eqref{equ:LMI_dd}.
%Hence, by satisfying \eqref{inequ:LMI_dd}, the closed-loop system is  exponentially differentially dissipative (as described by \eqref{eq:diff_diss}).
\end{proof}

\section{DCCM Synthesis and Controller Implementation}\label{sec:synthesis}
Based on the contraction \eqref{inequ:LMI} and differential dissipativity \eqref{equ:LMI_dd} conditions presented in Section \ref{sec:tractable},  here we develop control synthesis approaches from these conditions using Sum of squares (SOS) programming  (see \cite{wei2021discrete} for an alternative neural network-based approach). Details for numerical implementation of a contraction-based controller are also provided. 

\subsection{Overview of SOS Programming as a Synthesis Method}
SOS programming (see, e.g. \cite{boyd2004convex}) was proposed as a tractable method in \cite{manchester2017control} for computing CCMs for continuous-time control-affine nonlinear systems and is demonstrated in this work as an additionally tractable approach in the discrete-time setting. A polynomial $p(x)$, is an SOS polynomial, provided it satisfies
\begin{equation}\label{equ:SOS definition}
    p(x) = \sum_{i=1}^n{q_i(x)^2},
\end{equation} where $q_i(x)$ is a polynomial of $x$. Thus, it is easy to see that any SOS polynomial, $p$, is positive provided it can be expressed as in \eqref{equ:SOS definition}. Furthermore, in \cite{aylward2006algorithmic}, determining the SOS property expressed in \eqref{equ:SOS definition} is equivalent to finding a positive semi-definite $Q$ such that 
\begin{equation}
    p(x)=\phi(x)^\top Q\phi(x) \in \Sigma (x),
\end{equation}where $\phi(x)$ is a vector of monomials that is less or equal to half of the degree of polynomial $p(x)$, $\Sigma$ is stated in Notation. An SOS programming problem is an optimisation problem to find the decision variable $p(x)$ such that $p(x)\in\Sigma (x)$, which can be relaxed (in terms of computational complexity) for a particular range (e.g., a region of interest). Suppose we want $p(x)\in\Sigma (x)$ for $x \in [x_{i_{min}},x_{i_{max}}]$, with  $i =1,\cdots,n$. The relaxed condition can be constructed as follows,
\begin{equation}\label{equ:sos constraint}
\begin{aligned}
    &q(x) = p(x) - \sum_i^n{-(x_i-x_{i_{min}})(x_i-x_{i_{max}})c_i(x)} \in \Sigma (x)\\
    &\forall i, c_i(x) \in \Sigma (x),
\end{aligned}
\end{equation}
where the term $-(x_i-x_{i_{min}})(x_i-x_{i_{max}})c_i(x)$ represents a function that is only positive within the range $[x_{i_{min}},x_{i_{max}}]$. This means $p(x)$ only needs to be positive inside the constraints and $c_i(x)$ can be selected to relax the problem. In~\cite{Parrilo00}, the non-negativity of a polynomial is determined by solving  a semi-definite programming (SDP) problem (e.g. SeDuMi~\cite{Sturm1999}).
\subsection{DCCM Synthesis via SOS Programming}
Here we introduce SOS programming with relaxations for Theorem \ref{theorem:incremental stability}, followed by some discussion and natural progression to SOS programming for Proposition \ref{thm:condition}, and hence extensions for Proposition \ref{prop:dif dis}. From Theorem \ref{theorem:incremental stability}, two conditions need to be satisfied: the contraction condition \eqref{inequ: control affine contraction condition} and positive definite property of the matrix function $M$. These conditions can be transformed into an SOS programming problem (see, e.g.,~\cite{Parrilo00,aylward2006algorithmic}) if we assume the functions are all polynomial functions or polynomial approximations (see, e.g.,~\cite{EbeAll06}), i.e. 
\begin{equation}
    \begin{aligned}
                & \min_{k_c,m_c} \trace(M)\\
        s.t. ~ &\phi^\top  \Omega_1 \phi  \in \Sigma (x_k,u_k,\phi),\\
                &\phi^\top  M_k \phi \in \Sigma (x_k,\phi),
    \end{aligned}
\end{equation}where $\Omega_1 = -((A_k+B_kK_k)^\top M_{k+1}(A_k+B_kK_k) - (1-\beta)M_{k})$ represents the discrete-time contraction condition and $k_c,m_c$ are coefficients of polynomials for the controller gain, $K$ in \eqref{equ:differential feedback}, and metric, $M$, respectively (see the example in Section \ref{sec:example} for additional details). This programming problem is computationally difficult (if not intractable) due to the hard constraints imposed by the inequality \eqref{inequ: control affine contraction condition}. One possible improvement can be made by introducing relaxation parameters to soften the constraints (see, e.g. \cite{boyd2004convex}), i.e. introducing two small positive values, $r_1$ and $r_2$, as
\begin{equation}
    \begin{aligned}
        & \min_{k_c,m_c,r_1,r_2} r_1 + r_2 \\
        & s.t. ~ \phi^\top  \Omega_1 \phi - r_1I \in \Sigma (x_k,u_k,\phi)\\
        & \phi^\top  M_k \phi -r_2I \in \Sigma (x_k,\phi),r_1 \geq 0,r_2 \geq 0.
    \end{aligned}
\end{equation}
Note that the contraction condition holds if the two relaxation parameters $r_1$ and $r_2$ are some positive value, then we get a required DCCM as long as the relaxation parameters are positive. Although this relaxation reduces the programming problem difficulty, the problem remains infeasible, due to the terms coupled with unknowns, e.g. $B_kK_k^\top M_{k+1}B_kK_k$. 

Naturally, substitution for the equivalent contraction condition \eqref{inequ:LMI} in Proposition \eqref{thm:condition}, solves this computational obstacle, and hence a tractable SOS programming problem can be formed as follows
\begin{equation}\label{min:sos}
    \begin{aligned}
        &\underset{l_c,w_c,r}{\min} r\\
        &s.t. \ \tilde{\phi}^\top  \Omega \tilde{\phi} -rI \in \Sigma(x_k,u_k,\tilde{\phi}), ~r \geq 0,
    \end{aligned}
\end{equation} 
where $\Omega =         
\begin{bmatrix}
    W_{k+1}              & A_kW_k+B_kL_k \\
    (A_kW_k+B_kL_k)^\top     & (1-\beta)W_k
\end{bmatrix}$ and $w_c,l_c$ are the polynomial coefficients of the metric-controller dual $(W_k,L_k)$ (see the example in Section \ref{sec:example} for additional details). Note that the inverse and coupling terms are not present in \eqref{min:sos} and that its SDP tractable solution yields the matrix function $W_k$ and $L_k$, as required for contraction analysis and control.

%Finally, the controller \eqref{equ:control integral} can be constructed, following online computation of the geodesic \eqref{equ:geodesic} using the metric $M_k = W_k^{-1}$ and feedback gain $K_k = L_kW_k^{-1}$, as detailed in the following section.
%\subsection{Exponential Differential Dissipative DCCM Synthesis}
As an extension of this approach, synthesis of a DCCM and differential controller with desirable differential dissipativity properties, satisfying Proposition \ref{prop:dif dis} (and hence Theorem \ref{thm:dif dis}), can be expressed as the following SOS programming problem
\begin{equation}\label{min:sos dd}
    \begin{aligned}
        &\underset{l_c,w_c,r}{\min} r\\
        &s.t. \ \breve{\phi}^\top  \Phi \breve{\phi} -rI \in \Sigma(x_k,u_k,\nu_k,\breve{\phi}), ~r \geq 0,
    \end{aligned}
\end{equation} 
where $\Phi = \begin{bmatrix} 
W_{k+1} & A W_k + B_c L_k & B_\nu & 0\\
(A W_k + B_c L_k)^\top & (1 - \beta)W_k & W_k S & W_k \\ 
B_\nu^\top & S^\top W_k & R & 0\\
0 & W_k & 0 & -Q^{-1} 
\end{bmatrix}$. 

\subsection{Numerical Implementation}
Suppose that the optimisation problem \eqref{min:sos} (or \eqref{min:sos dd}) is solved for the pair $(W,L)$ and hence $(M,K)$ are obtained. The next step in implementing the contraction-based controller \eqref{equ:control integral} is to integrate \eqref{equ:differential feedback} along the geodesic, $\gamma$ \eqref{equ:geodesic}. 
Subsequently, one method to numerically approximate the geodesic is shown. From \eqref{equ:Riemannian distance and energy} and \eqref{equ:geodesic}, we have the following expression for computing the geodesic, 
\begin{equation}\label{equ:geo cal}
    \gamma(x,x^*) = \argmin_c \int_0^1{\frac{\partial c(s)}{\partial s}^\top  M(c(s)) \frac{\partial c(s)}{\partial s}ds}.
\end{equation}
%The form is not explicit in parameter $s$ as the function also depends on path states $c(s)$ which needs the information of the geodesic. Thus it cannot be solved analytically. That is why a numerical method is required. 
where (see Section \ref{sec:discrete-time contraction analysis}) $c(s)$ is an $s$-parameterized smooth curve connecting $x$ ($s=0$) to $x^*$ ($s=1$). Since \eqref{equ:geo cal} is an infinite  dimensional problem over all smooth curves, without explicit analytical solution, the problem must be discretized to be numerically solved. Note that the integral can be approximated by discrete summation provided the discrete steps are sufficiently small. As a result, the geodesic \eqref{equ:geo cal} can be numerically calculated by solving the following optimization problem,
% \begin{equation}\label{min:geodesic}
%     \begin{aligned}
%     \bar \gamma(x_0,x_1) = \argmin_{\Delta x_{s}} &\sum_{i=1}^N{\Delta x_{s_i}^\top  M(x_i) \Delta x_{s_i} \Delta s_i}&\\
%     s.t.  &\sum_{i=1}^N \Delta x_{s_i} \Delta s_i = x_1 - x_0&\\
%           &\sum_{i=1}^N \Delta s_i = 1,\quad  x_i = \sum_{j=1}^i \Delta x_{s_j} \Delta s_j + x_0
%     \end{aligned}
% \end{equation}
\begin{equation}\label{min:geodesic}
\begin{aligned}
    \bar \gamma(x,x^*) = \argmin_{\Delta \tilde{x}_{s}} &\sum_{i=1}^{N}{\Delta \tilde{x}_{s_i}^\top  M_{NN}(\tilde{x}_i) \Delta \tilde{x}_{s_i} \Delta s_i}\\
    s.t. \quad  & \tilde{x}_1 = x,~\tilde{x}_N = x^*,
\end{aligned}
\end{equation}where $\bar \gamma(x,x^*) \approx \gamma(x,x^*)$ represents the numerically approximated geodesic, $x$ and $x^*$ are the endpoints of the geodesic, $\tilde{x}_i$ represents $i$-th point on a discrete path in the state space, $\Delta \tilde{x}_{s_i} := \Delta {\tilde{x}_i} / \Delta {s_i} \approx {\partial c(s)}/{\partial s}$ can be interpreted as the displacement vector discretized with respect to the $s$ parameter, $\Delta \tilde{x}_s:=(\Delta \tilde{x}_{s_1},\cdots,\Delta\tilde{x}_{s_N})$ is the discretized path joining $x$ to $x^*$ (i.e., discretization of c(s) in~\eqref{equ:geo cal}), all $\Delta {s_i}$ are small positive scalar values chosen such that $\sum_{i=1}^N \Delta s_i = 1$, $N$ is the chosen number of discretization steps (of s), $\tilde{x}_i = \sum_{j=1}^i \Delta \tilde{x}_{s_j} \Delta s_j + x$ represents the numerical state evaluation along the geodesic.

Note that \eqref{min:geodesic} is the discretization of \eqref{equ:geo cal} with $\Delta \tilde{x}_{s_i}$ and $\Delta_{s_i}$ as the discretizations of  $\frac{\partial c(s)}{\partial s}$ and $\delta_s$ respectively, whereby the constraints in \eqref{min:geodesic} ensure that the discretized path connecting the start, $x$, and end, $x^*$, state values align with the continuous integral from $s=0$ to $s=1$. Hence, as $\Delta_{s_i}$ approaches 0, i.e., for an infinitesimally small discretization step size, the approximated discrete summation in \eqref{min:geodesic} converges to the smooth integral in \eqref{equ:geo cal}.

After the geodesic is numerically calculated using \eqref{min:geodesic}, the control law in \eqref{equ:control integral} can be analogously calculated using an equivalent discretisation as follows
\begin{equation}\label{equ:controller}
    u_k = u_k^* +  \sum_{i=1}^N K_k(\tilde{x}_i) \Delta \tilde{x}_{s_i} \Delta s_i.
\end{equation}
Substituting $K_k = L_kW_k^{-1}$ into \eqref{equ:controller}, we can then implement the control law as follows
\begin{equation}
\label{equ:disccontr}
    u_k = u_k^* +  \sum_{i=1}^N L_k(\tilde{x}_i)W_k^{-1}(\tilde{x}_i) \Delta \tilde{x}_{s_i} \Delta s_i,
\end{equation}
%\begin{equation}\label{equ:controller}
%    u_k = u_k^* +  \sum_{i=1}^N \Delta x_{s_i} \Delta s_i K(x_i).
%\end{equation}
where $(x_k^*,u_k^*)$ are the state and control reference trajectories at time $k$. Note that this real-time control approach results in a setpoint-independent control structure, whereby the structure (imposed  by matrix functions $L$ and $W$) required to implement this control law are obtained offline via solution to \eqref{min:geodesic}. To see this, recall that $\tilde{x}_i$ represents the numerical state evaluation along the geodesic at discrete points and $K(\tilde{x}_i)$ is the evaluation of the matrix function $K=LW^{-1}$ at each of those points. Then, as the desired state value, $x_k^*$, changes, the feed-forward component, $u_k^*$, can be instantly updated, and the feedback component, $\sum L_k(\tilde{x}_i)W_k^{-1}(\tilde{x}_i) \Delta \tilde{x}_{s_i} \Delta s_i$ (i.e., $\int K(\gamma)\delta_\gamma ds$ in \eqref{equ:control integral}), can be automatically updated through online geodesic calculation.

\section{Illustrative Example}\label{sec:example}
In this section, we present a case study to illustrate the synthesis and implementation of a contraction-based controller using the proposed approach. \sloppy Consider a well mixed, nonisothermal CSTR where 3 parallel irreversible elementary exothermic reactions take place of the form $A \to B$, $A \to C$, $A \to D$ \cite{christofides2011networked}, whereby the process can be modelled as follows,
\begin{equation}\label{equ:sim sys}
\begin{aligned}
    x_{1_{k+1}} &= x_{1_k} + \Delta_t \left(\frac{F}{V_r}(C_{A0} -x_{1_k}) + \nu_k +  \sum_{i=1}^3 \kappa_{i0} e^{\frac{-E_i}{R x_{2_k}}} x_{1_k}\right) \\
    x_{2_{k+1}} &= x_{2_k} + \Delta_t \left(\frac{F}{V_r}(T_{A0}-x_{2_k}) - \sum_{i=1}^3 \frac{\Delta H_i}{\sigma c_p} k_{i0} e^{\frac{-E_i}{R x_{2_k}}}x_{1_k} + u_k \right),
\end{aligned}
\end{equation}
% \begin{strip}
% \begin{align}
%     C_A(k+1) &= C_A(k) + \Delta_t (\frac{F}{V_r}(C_{A0} -C_A(k)) + \sum_{i=1}^3 k_{i0} e^{\frac{-E_i}{R T}} C_A(k) + \nu(k)) \\
%     T(k+1) &= T(k) + \Delta_t (\frac{F}{V_r}(T_{A0}-T(k)) - \sum_{i=1}^3 \frac{\Delta H_i}{\sigma c_p} k_{i0} e^{\frac{-E_i}{R T}}C_A(k) + u(k)),
% \end{align}
% \end{strip}
where the feed to the reactor consists of reactant $A$ with molar concentration $C_{A0}$ at flow rate $F$ and temperature $T_{A0}$. This process model consists of two states, where $x_1 = C_A$ denotes the concentration of reactant $A$, and $x_2 = T$ denotes the temperature of the reactor. The states $x_1$ and $x_2$ are controlled by manipulating  $u = \frac{Q}{\sigma C_p V_r}$, where $Q$ represents the rate of heat input/removal. The process disturbance $\nu$ is caused by the variation in the feed concentration $\Delta C_{A0}$ with $\nu =\frac{F}{V_r} \Delta C_{A0}$. For the remaining process variables, $V_r$ denotes the volume of the reactor, $\Delta H_i,\kappa_{i0},E_i$ denotes the enthalpies, preexponential constants and activation energies of the three reactions, respectively, and $c_p$ and $\sigma$ denote the heat capacity and density of the fluid in the reactor, respectively. For simulation purposes, the system model is normalized in the range of operation with a sampling period of $\Delta_t = 0.05\ h$. All parameters are shown in Table \ref{tab:process parameters}.
\begin{table}[]
    \centering
    \begin{tabular}{cc|cc}
        $F$          & $4.998[m^3/h]$               & $\kappa_{10}$ & $3\times 10^6[h^{-1}]$\\
        $V_r$        & $1[m^3]$                     & $\kappa_{20}$ & $3\times 10^5[h^{-1}]$\\
        $R$          & $8.314[KJ/kmol K]$           & $\kappa_{30}$ & $3\times 10^5[h^{-1}]$\\
        $T_{A0}$     & $300[K]$                     & $E_1$    & $5\times 10^4[KJ/kmol]$    \\
        $C_{A0}$     & $4[kmol/m^3]$                & $E_2$    & $7.53\times 10^4[KJ/kmol]$ \\
        $\Delta H_1$ & $-5.0 \times 10^4[KJ/kmol]$  & $E_3$    & $7.53\times 10^4[KJ/kmol]$ \\
        $\Delta H_2$ & $-5.2 \times 10^4[KJ/kmol]$  & $\sigma$ & $1000[kg/m^3]$             \\
        $\Delta H_3$ & $-5.4 \times 10^4[KJ/kmol]$  & $c_p$    & $0.231[KJ/kg K]$             
    \end{tabular}
    \caption{Process Parameters}
    \label{tab:process parameters}
\end{table}

To synthesise a metric and controller pair $(M,K)$, the exponential functions in \eqref{equ:sim sys} are first approximated using second order polynomial functions, such that the system description is amenable to SOS programming. Then, the Jacobian matrices, $A_{k},B_{c,k},B_{\nu,k}$, are calculated to obtain the corresponding differential system model \eqref{eq:sys_d diff}. As required for the differential dissipativity-based SOS problem \eqref{min:sos dd}, a $(Q,S,R)-$supply rate, with $Q=I$, $S=0$, $R=0.81I$, is selected to achieve a truncated incremental $\mathcal{L}_2$ gain of $0.9$ (from the disturbance to state). The respective contraction metric and feedback gain duals, $W_k$ and $L_k$, are matrices of polynomial functions in the following forms, i.e.
\begin{equation}
\label{equ:WkLk}
    W_k = 
    \begin{bmatrix}
        W_{{11}_k} & W_{{12}_k} \\
        W_{{12}_k} & W_{{22}_k}
    \end{bmatrix}, \qquad
    L_k =
    \begin{bmatrix}
        L_{1_k} \\
        L_{2_k}
    \end{bmatrix},
\end{equation}
where $W_{{\cdot \cdot}_k}=w_{{\cdot \cdot c}}v(x_k)$ and $w_{{\cdot \cdot c}}$ is a row vector of unknown coefficients, and similarly for $L_k$. As required to solve the SOS problem in \eqref{min:sos} and \eqref{min:sos dd}, the functions $W_{{\cdot \cdot k}}$  need to be polynomial functions and as such are expressed using the common monomial vector, $\phi(x_k)$, defined as 
\begin{equation}
    \phi(x_k) = 
    \begin{bmatrix}
        x_{1_k}^4 & x_{1_k}^3x_{2_k} & \cdots & x_{2_k}^4
    \end{bmatrix}^\top ,
\end{equation}
where the polynomial order is chosen to be 4. 
%\begin{equation}
%    L_k =
%    \begin{bmatrix}
%        L_{1_k} \\
%        L_{2_k}
%    \end{bmatrix},
%\end{equation}where $L_{1_k} = l_{1c}V(x_k)$ with $l_{1c}$ be the unknown coefficient row vector and same for $l_2$. 
Additionally, $W_{k+1}$, is constructed as a matrix of polynomials
\begin{equation}
    W_{k+1} = 
    \begin{bmatrix}
        W_{{11}_{k+1}} & W_{{12}_{k+1}} \\
        W_{{12}_{k+1}} & W_{{22}_{k+1}}
    \end{bmatrix},
\end{equation}
where the elements are defined as $W_{{\cdot \cdot}_{k+1}}=w_{{\cdot\cdot c}}\phi(x_{k+1})$ and $w_{{\cdot \cdot c}}$ is the same coefficient vector for $W_k$ in \eqref{equ:WkLk}. Two design scenarios are then considered: i) without dissipativity, i.e. solving the optimisation problem in \eqref{min:sos}, and; ii) with dissipativity, i.e., solving the optimisation problem in \eqref{min:sos dd}. Both SOS problems were then solved for the contraction rate, $\beta=0.9$, and by considering the normalised range of operation $x_{min} = 0.01,x_{max} = 1$, with the resulting metric and controller matrix coefficients shown in \ref{Appen}. 

A time-varying reference, corresponding to the variations in product specifications and energy cost, was considered as the sequence of setpoints \sloppy $(x_1^*,x_2^*,u^*)$=$(0.750,0.680,2.241)$, $(0.850,0.716,2.264)$, $(0.950,0.748,2.269)$ on the respective $k \Delta_t$ intervals of  $[0,10.0)$, $[10.0,19.9)$, $[19.9,29.8]$ $h$. Such feasible setpoints can be generated using an RTO layer -- a befittingly popular method for real-time process control. It is worth noting that for contraction-based control the optimal or even \textit{complete} trajectories (e.g., between the current state and target setpoint) are not needed, only the desired feasible setpoints are required. Suppose that a feasible complete set of reference trajectories was in fact available (typically a non-trivial task), then the same contraction based controller could be additionally used to drive the system to such references, and without structural redesign (simply update the geodesic information in \eqref{equ:control integral}). The contraction-based control methods does not require redesigning the control algorithm as the reference changes (setpoint or otherwise), unlike the Lyapunov-based control designs. 

\begin{figure}
    \begin{center}
        \includegraphics[width=\linewidth]{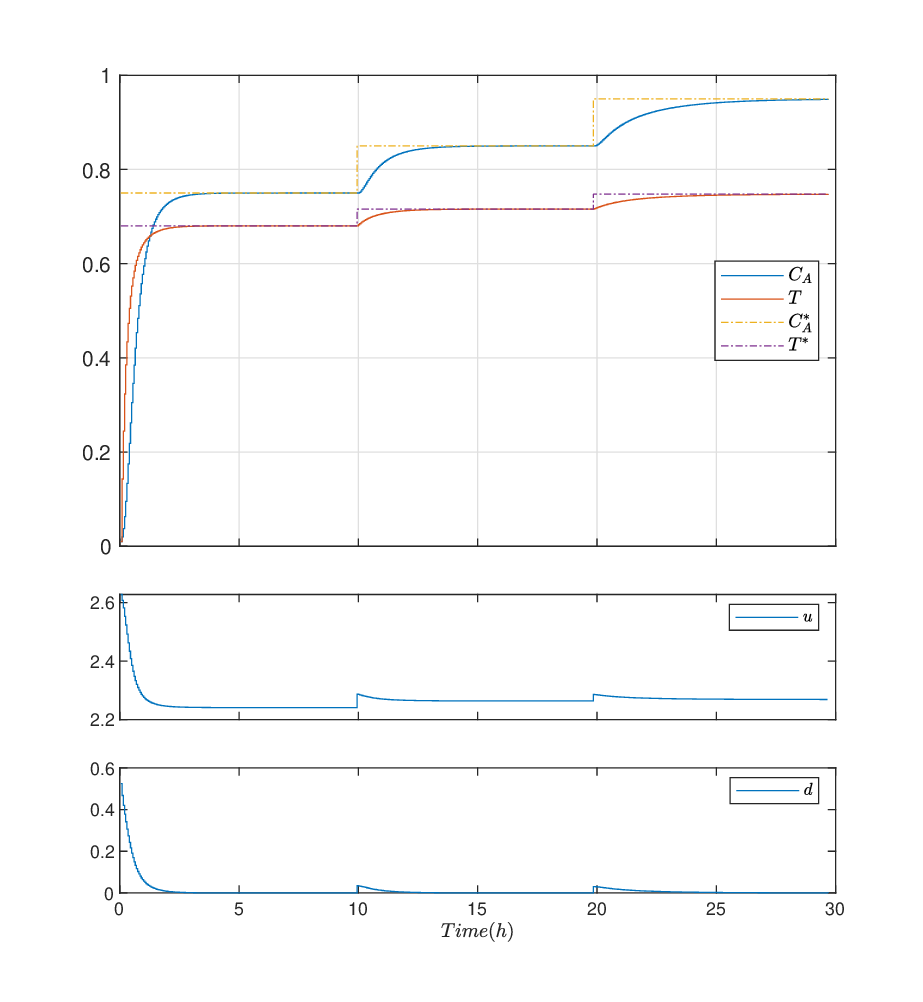}
        \caption{Contraction-based control of the CSTR \eqref{equ:sim sys} without disturbance}
        \label{fig:cstr sim}
    \end{center}
\end{figure}

\begin{figure}
    \begin{center}
        \includegraphics[width=\linewidth]{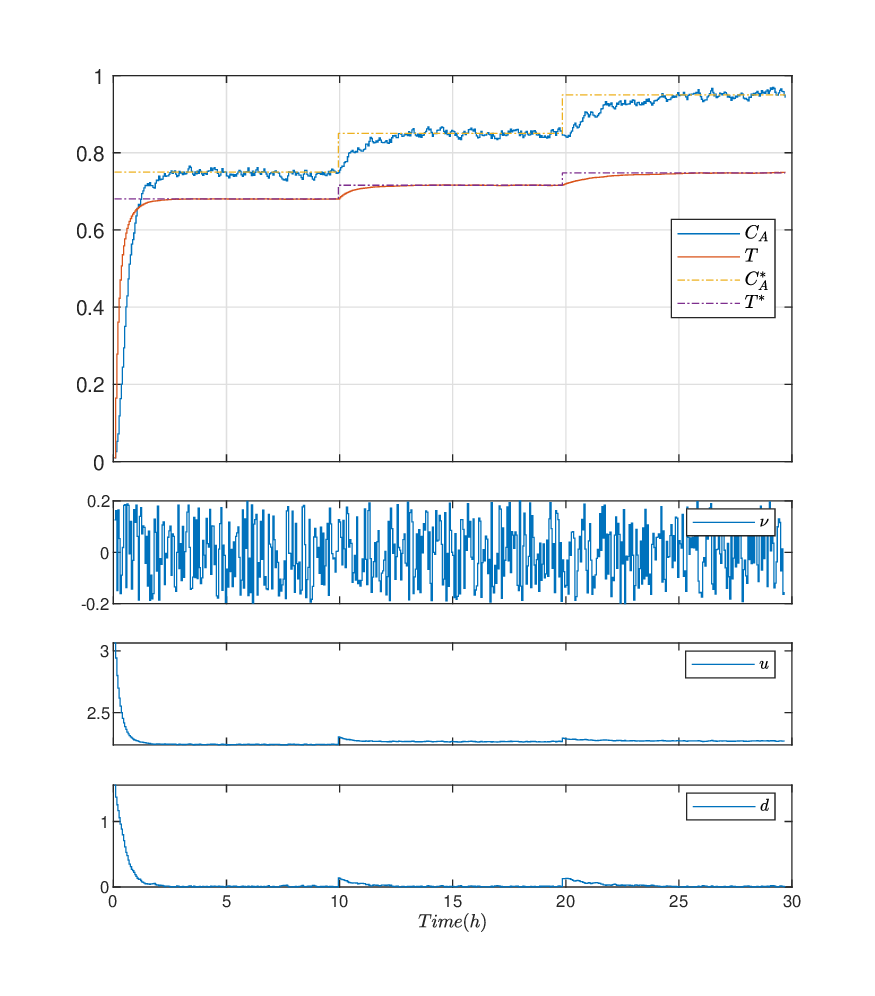}
        \caption{Contraction-based control of the CSTR \eqref{equ:sim sys} with disturbance attenuation}
        \label{fig:cstr sim d}
    \end{center}
\end{figure}

The CSTR \eqref{equ:sim sys} was then simulated with the discrete-time contraction-based controller \eqref{equ:disccontr} under two scenarios: i) without disturbance, see Figure \ref{fig:cstr sim}, and; ii) with feed concentration disturbance, see Figure \ref{fig:cstr sim d}. The results shown in Figure \ref{fig:cstr sim} demonstrate that the proposed  contraction-based controller is capable of driving the CSTR error-free to the desired time-varying product specifications in the absence of disturbances (as per Theorem \ref{theorem:incremental stability}). Figure \ref{fig:cstr sim d} shows that the contraction-based design which incorporates a differential dissipativity condition additionally achieves disturbance attenuation (as per Theorem \ref{thm:dif dis}). 

% \begin{equation}
% \begin{aligned}
%     \frac{d x_A}{dt} &= \frac{F_0}{V}(x_{A0}-x_A) + \frac{F_r}{V}(x_{Ar}-x_A) - k_{AB} e^{\frac{-E}{R T}} x_A\\
%     \frac{d x_B}{dt} &= \frac{F_0}{V}(x_{B0}-x_B) + \frac{F_r}{V}(x_{Br}-x_B) + k_{AB} e^{\frac{-E}{R T}} x_A - k_{BC} e^{\frac{-E}{R T}} x_B\\
%     \frac{d T}{dt} &= \frac{F_0}{V}(T_0-T) + \frac{-\Delta H_{AB}}{C_p}k_{AB} e^{\frac{-E}{R T}}x_A + \frac{-\Delta H_{BC}}{C_p}k_{BC} e^{\frac{-E}{R T}}x_B + \frac{Q}{\rho C_p V}
% \end{aligned}
% \end{equation}
% \begin{equation}
% \begin{aligned}
%     x_{Ar} &= \frac{\alpha_A x_A}{(\alpha_A-\alpha_C) x_A + (\alpha_B-\alpha_C) x_B + \alpha_C}\\
%     x_{Br} &= \frac{\alpha_B x_B}{(\alpha_A-\alpha_C) x_A + (\alpha_B-\alpha_C) x_B + \alpha_C}
% \end{aligned}
% \end{equation}

\section{Conclusion}\label{sec:conclusion}
A systematic approach to the implementation of contraction-based control for discrete-time nonlinear processes was developed. Through the differential system framework, contraction and dissipativity conditions were derived, to ensure stability of the resulting closed-loop control system, i.e., exponential convergence to feasible time-varying references and bounded disturbance response. Computationally tractable equivalent conditions were then derived and additionally transformed into SOS programming problems, such that joint synthesis of a discrete-time control contraction metric and stabilising feedback controller could be completed with desired closed-loop properties. Complete controller implementation details were provided and the overall approach was effectively demonstrated through a numerical CSTR case study.



%% References


% \bibliographystyle{elsarticle-num}
% \bibliography{<your-bib-database>}
\bibliographystyle{elsarticle-num}
\bibliography{Wei_McCloy_Bao}





% \end{thebibliography}
\appendix
\section{Coefficients of Matrix Functions in Section~\ref{sec:example}}\label{Appen}
\label{app:coeff}
The polynomial coefficient vectors of $W$ and $L$ used in the illustrative CSTR example of Section \ref{sec:example} are as follows.

Without disturbance consideration (i.e., solution to \eqref{min:sos}):

$w_{{11c}}$ =
   [3.6663
    1.9921
   -0.0538
    2.0161
   -0.0442
    1.7564
    1.1821
    0.1662
    0.9565
   -0.7217
    3.1153
    0.0425
    2.8042
    0.0970
    3.9131],
    
$w_{{12c}}$ =
    [-0.3123
    0.0911
   -0.4714
    0.2899
   -0.1363
   -0.3873
    0.3568
   -0.1626
   -0.1961
   -0.7237
    0.0694
    0.0653
    0.0116
    0.0367
   -0.2880],
   
$w_{{22c}}$ =
    [1.8674
    1.3870
   -0.8890
    1.5241
   -0.2636
    0.4206
    0.9464
    0.0930
    0.8197
   -1.8968
    2.5797
    0.1082
    2.5884
    0.1910
    2.1615],
 
     $l_{1c}$ =
   [-0.0531
   -0.0365
   -0.0217
   -0.0012
   -0.0399
   -0.0053
   $-0.0050$
    0.0014
   -0.0392
    0.0075
   -0.0015
   -0.0051
    0.0032
   $-0.0418$
    0.0159],

     $l_{2c}$ =
    [-1.1214
   -0.0260
   -1.0183
   -0.0518
   -0.0144
   -1.0058
   -0.0078
   -0.0358
   -0.0099
   -1.0294
   -0.0412
   -0.0029
   -0.0364
   $-0.0071$
   -1.1031].

With disturbance attenuation (i.e., solution to \eqref{min:sos dd}):

$w_{{11c}}$ =
   [0.6289
    0.0045
   -0.3167
    0.2972
   -0.0015
    0.0220
   $-0.0082$
   -0.0899
   -0.0025
   -0.5972
    0.0905
    0.0018
    0.5744
   $-0.0003$
    0.6306],
    
$w_{{12c}}$ =
    [-0.0661
   -0.0008
   -0.0312
   -0.0073
    0.0004
    0.0541
   $-0.0001$
   -0.0346
    0.0002
   -0.0395
    0.0006
    0.0002
   -0.0106
    0.0005
   -0.0232],
   
$w_{{22c}}$ =
    [0.4219
    0.0035
   -0.3036
    0.2731
   -0.0027
    0.0444
   $-0.0075$
   -0.1255
   -0.0012
   -0.5799
    0.0892
    0.0018
    0.5190
    0.0002
    0.5106],
 
     $l_{1c}$ =
   [0.1181
   -0.0000
    0.1272
    0.0003
   -0.0000
    0.1361
    0.0000
    0.0003
   -0.0000
    0.1450
    0.0001
    0.0000
    0.0003
   -0.0000
    0.1538],

     $l_{2c}$ =
    [-0.3181
   -0.0002
   -0.3036
   -0.0065
   -0.0001
   -0.3039
    0.0000
   -0.0039
   -0.0001
   -0.3099
   -0.0021
    0.0000
   -0.0040
   $-0.0000$
   -0.3234].
\end{document}